\documentclass[11pt]{article}

\usepackage{amsmath,amssymb,amsthm,graphicx,latexsym}
\newcommand{\ed}{\end{document}}
\newcommand{\beq}{\begin{equation}}
\newcommand{\eeq}{\end{equation}}
\newcommand{\beqa}{\begin{eqnarray}}
\newcommand{\eeqa}{\end{eqnarray}}
\newcommand{\bc}{\begin{center}}
\newcommand{\ec}{\end{center}}

\newcommand{\ba}{\begin{array}}
\newcommand{\ea}{\end{array}}

\begin{document}
%%%%%%%%%%%%%%%%%%%%%%%%%%%%%%%%%%%%%%%%%%%
%%%%%%%%%%%%%%%%%%%%%%%%%%%%%%%%%%%%%
\title{{\bf{Hamiltonian analysis of interacting fluids }}}
%%%%%%%%%%%%%%%%%%%%
\author{  {\bf {\normalsize Rabin Banerjee$^1$}$
$\thanks{E-mail: rabin@bose.res.in}},$~$
{\bf {\normalsize Subir Ghosh$^2$}$
$\thanks{E-mail: subir\_ ghosh2@rediffmail.com}},$~$
{\bf {\normalsize Arpan Krishna Mitra$^1$}$
$\thanks{E-mail: arpan@bose.res.in}}$~$
\\{\normalsize $^1$S. N. Bose National Centre for Basic Sciences,}
\\{\normalsize JD Block, Sector III, Salt Lake, Kolkata-700098, India}
\\\\
{\normalsize $^2$Indian Statistical Institute}
\\{\normalsize  203, Barrackpore Trunk Road, Kolkata 700108, India}
\\[0.3cm]
}
\date{}
\maketitle

\begin{abstract}
Ideal fluid dynamics is studied as a  relativistic field theory with particular importance on its hamiltonian structure. The Schwinger condition, whose integrated version yields the stress tensor conservation, is explicitly verified both in equal-time and light-cone coordinate systems. We also consider the hamiltonian formulation of fluids interacting with an external gauge field. The complementary roles of the canonical(Noether) stress tensor and the symmetric one obtained by metric variation are discussed. 
\end{abstract}
\pagebreak
\section {Introduction}
Fluid dynamics as an applied science has a long history but its generalization as a relativistic theory and its subsequent analysis as a relativistic field theory is a relatively recent development. The Euler formulation of the fluid system in terms of the density $\rho (x)$ and velocity fields $v_i(x)$ (in a non-relativistic framework) is suitable for this purpose \cite{lan, hol, mor}. The hydrodynamic equations are essentially the local conservation laws supplemented by the constitutive relations that express the stress tensor in terms of the fluid variables. These notions are extended to the relativistic case by introducing a comoving velocity $u_{\mu}$ normalised as $u^{\mu}u_{\mu}=1$.

A lagrangian version of fluid dynamics is plagued with obstructions  due to the presence of a Casimir operator, the vortex helicity (see Jackiw et al. \cite{baz, jac} for a modern perspective). The problem can be cured by the introduction of Clebsch variables \cite{cl, kara} designed in such a way that the  vortex helicity becomes a surface contribution and does not obstruct the lagrangian formulation. Extension of these ideas in a relativistic context has also been dealt with. But all these studies are concerned with a free (or at best self-interacting) fluid and an in depth hamiltonian analysis of a relativistic fluid with external gauge interactions remains unexplored.

Let us elaborate on our work from this perspective. We have   presented a systematic and detailed analysis of an ideal relativistic fluid in the hamiltonian framework. Subsequently this analysis is generalised to include interaction with an external gauge field. Previous works in this direction are \cite{sch, sc, koda}. Introduction of the Clebsch variables  reduces the system to a first order one: a constraint system in the Dirac formalism \cite{dir} (see also \cite{hrt}). We study both the systems in Dirac's framework. The relevant constraints are identified and the systems are found to be second class. The modified symplectic structure is the same in both cases. Our analysis reveals that the relativistic Eulerian fluid model poses an intriguing example of a Hamiltonian constraint system. This becomes manifest especially when gauge interactions are taken into account.

The crux of the problem is the construction of the stress tensor. There are two conventional formalisms for deriving the stress tensor. The canonical $T_{\mu\nu}$ is obtained via Noether prescription and the symmetric  $\Theta_{\mu\nu}$ is obtained by metric variation. For the free theory both definitions agree. However in the presence of interaction, $T_{\mu\nu}$ and  $\Theta_{\mu\nu}$ do not match. The former generates the correct equations of motion for all the dynamical variables but does not yield the correct conservation law of the stress tensor. The latter, on the other hand, satisfies the correct conservation law but it fails to generate the correct equation of motion for one of the fluid variables. At the same time we show an interesting connection between  $T_{\mu\nu}$ and  $\Theta_{\mu\nu}$. A simple modification of $T_{\mu\nu}$ yields the correct conservation law. Furthermore, the modified version is shown to be identical to $\Theta_{\mu \nu}$. This provides an internal consistency. We stress that  these  are new observations that were  not revealed in the literature that dealt with fluid models.

We have made a detailed study of the Schwinger condition both in equal time and light-cone coordinates. Its role in conservation laws on which the dynamics of fluids is based is discussed. The fact that the Schwinger condition holds for classical fluids is a new observation.

It may be mentioned that the recent idea of fluid-gravity correspondence \cite{ranga, hu} has brought, to the forefront, the theoretical study of fluid dynamics from a high energy and gravitational physics perspective. The basic premise is that relativistic or non-relativistic fluid dynamics can reproduce the low energy behavior of systems in local thermal equilibrium in a universal way. Indeed,
this is an offshoot of the AdS/CFT correspondence \cite{2} that paves the way for studying strongly coupled systems from their weakly coupled analogues in one dimension higher. Generically one exploits AdS/CFT correspondence to study strongly correlated condensed matter systems as boundary conformal theories from results obtained in weakly coupled classical gravity theories in one higher dimension. However, the mutual exchange of ideas can work bothways in fluid-gravity correspondence: fluid systems can yield results relevant in eg. black hole physics, Hawking radiation \cite{unruh}  while
gravitational physics can provide new ideas in the context of viscous fluids, turbulence, to name a few. All these considerations require a systematic study of the fluid system as a field theory in the Euler scheme, which is essentially a hamiltonian framework. Our analysis in this paper is geared towards this direction providing some new results and fresh insights.

The paper is organized as follows: In section 2 the relativistic fluid model in terms of Clebsch variables is introduced in equal-time coordinate system. The sysmplectic structure is derived in the  hamiltonian formalism and the Schwinger condition is verified.  Section 3 deals with the fluid in light-cone coordinate system. The  dynamics and light-cone Schwinger condition are discussed. The interacting fluid system is analysed in section 4. The paper ends with our conclusions in section 5.

\vspace{1em}
\section{Relativistic fluid mechanics in equal-time coordinates}
We are going to describe the dynamics of a relativistic ideal fluid in this section. Usually this dynamics is espressed by the conservation of the stress tensor
\begin{equation}
\label{one}
\partial_{\mu}\Theta^{\mu\nu}=0
\end{equation}
which is further supplemented by the constitutive relation, 
\begin{equation}
\label{bl}
\Theta_{\mu \nu}= -\eta_{\mu \nu}P_{rel}+(\epsilon_{rel}+P_{rel})u_{\mu}u_{\nu}
\end{equation}
that gives the stress tensor in terms of the relativistic fluid variables, the pressure $P_{rel}$, the energy density $\epsilon_{rel}$ and the comoving velocity $u_{\mu}$ satisfying $u^{\mu}u_{\mu}=1$.

 However we start with a manifestly Lorentz covariant lagrangian density by introducing a generalized scalar potential function $f(\sqrt{j^{\mu}j_{\mu}})$ as for instance done by \cite{jac}. Here $j^{\mu}$ is the current Lorentz vector $j^{\mu}=(\rho, {\bf{j}})$ satisfying the continuity equation
 \begin{equation}
\label{e0}
\partial_\mu j^\mu =0
\end{equation}
so that if necessary one may couple it to background gauge field. The appropriate lagrangian density is given by
 \begin{equation}
\label{a}
 {\cal{L}}=-\eta^{\mu\nu}j_{\mu}a_{\nu}-f;~~ ~ \eta^{\mu\nu}=diag(1,-1,-1,-1)
 \end{equation}
 where $a_{\mu}$ is defined in terms of three scalar Clebsch variables  $\theta , \alpha, \beta $\cite{cl, kara},
\begin{equation}
  \label{b}
  a_{\mu}=\partial_{\mu}\theta+\alpha\partial_{\mu}\beta .
  \end{equation}
We shall subsequently show that the energy momentum tensor derived from this lagrangian density will satisfy (\ref{one}) and (\ref{bl}) while the current entering (\ref{a}) satisfies (\ref{e0}).

We take (\ref{a}) as the lagrangian density of an ideal relativistic fluid\cite{jac}. It is worthwhile to point out here a contrast between the Lagrangian (point particle) and Euler (field theoretic) frameworks of fluid mechanics. In the former one has constraints so that not all coordinates $x_\mu$ are independent whereas no such constraint is present in the latter. Since effectively the (Lagrangian) velocity is replaced by $a_\mu$, $a_\mu$ is explicitly written in terms of {\it{three}} (and not four) degrees of freedom $\theta , \alpha, \beta $. (For a discussion on this point see \cite{kara}.) Furthermore, the reason to introduce  Clebsch variables has also been discussed in the Introduction.

 The expanded form of the lagrangian (\ref{a}) with $j^{\mu}j_{\mu}=n^{2}$, is
 \begin{equation}
 \label{c}
 {\cal{L}}= -\rho\partial_{0}\theta-j^{i}\partial_{i}\theta-\rho\alpha\partial_{0}\beta-j^{i}\alpha\partial_{i}\beta-f(n).
 \end{equation}
 In the above  we have defined $\rho=j^{0}$.
 Our prescription is the following: the variables associated with time derivatives like
   $\rho,\alpha,\beta,\theta$ are treated as dynamical  whereas $j^{i}$ are regarded as  auxiliary variables. From the lagrangian (\ref{c}), equations obtained by varying $\beta, \alpha $, $\rho$ and $j_{\mu}$ are, respectively\footnote{Prime of a function indicates differentiation, thus $f'(n)=\frac{df(n)}{dn}$. },
      \begin{equation}
     \label{e1}
  j^\mu\partial _\mu\alpha =0,
  \end{equation}
     \begin{equation}
  \label{e2}
j^\mu\partial _\mu\beta =0,
  \end{equation}
     \begin{equation}
  \label{e3}
\dot{\theta}+ \alpha\dot{\beta}+\dfrac{\rho}{n}{f^{\prime }(n)}=0.
  \end{equation}
  \begin{equation}
\label{r}
j_{\mu}=-\frac{n}{f^{\prime}(n)}a_{\mu}=-\frac{n}{f^{\prime}(n)}(\partial_{\mu}\theta + \alpha \partial_{\mu}\beta).
\end{equation}

  Note that variation of $\theta$ reproduces the current conservation law (\ref{e0}).
We stress that the status of the last equation (\ref{r}) is distinct from the previous ones (\ref{e1}-\ref{e3}). Its time component is just(\ref{e3}). Now, (\ref{e1}, \ref{e2}, \ref{e3}) represent genuine equations of motion since these involve the velocities\footnote{For a second order system the true equations of motion involve the accelerations but for a first order system like(\ref{c}), these equations involve the velocities. }. The space component of (\ref{r}), on the contrary, is more like a constraint than an equation of motion since it is bereft of any velocity term. Not surprisingly this equation is obtained by varying $j_{i}$ which is regarded as an auxiliary variable. It needs to be interpreted carefully and a specific prescription is required (as we will provide later) for its application.

   Let us now develop a  hamiltonian formulation. Being first order in time derivatives the system is a constraint system and has a non-trivial symplectic structure, that can be identified with the  Dirac brackets of the variables in a hamiltonian formalism  \cite{dir}.    The first step is to define the conjugate momenta for the dynamical variables, which are
 \begin{equation}
 \label{d}
 \pi_{\theta}=\frac{\partial{\cal{L}}}{\partial\dot{\theta}}=- \rho;~ ~ \pi_{\alpha}=\frac{\partial{\cal{L}}}{\partial \dot{\alpha}}=0; ~~  \pi_{\beta}=\frac{\partial{\cal{L}}}{\partial \dot{\beta}}=-\rho \alpha, ~~ \pi_ {\rho}=\frac{\partial{\cal{L}}}{\partial \dot{\rho}}=0.
 \end{equation}
 They yield  four primary constraints
 \begin{equation}
 \label{e}
 \Omega_{1}=\pi_{\theta}+\rho \approx 0; ~~\Omega_{2}=\pi_{\alpha}\approx 0; ~~\Omega_{3}=\pi_{\beta}+\rho\alpha\approx 0; ~~\Omega_{4}=\pi_{\rho}\approx 0.
 \end{equation}
 Using canonical Poisson brackets of the generic form\footnote{Here $
	\bf{x}$ denotes space components $x_i$.} $\{q(x),\pi _q(y)\} =\delta (\bf {x}-\bf{y})$, we can easily show that the constraint algebra does not close indicating that they form a set of four second class constraints \cite{dir}.
 In a generic  system with $n$ second class constraints  $\Omega_i$, $i=1,2,..n$, the modified symplectic structure (or Dirac brackets) are defined in the following way,
\begin{equation}
\{A,B\}^*=\{A,B\}-\{A,\Omega _i\}\{\Omega ^i,\Omega ^j\}^{-1}\{\Omega _j,B\}, \label{a6}
\end{equation}
where $\{\Omega ^i,\Omega ^j\}$ is the invertible constraint matrix. From now on we will only use Dirac brackets but for notational simplicity we will refer to them as $\{,\}$  instead of $\{,\}^*$. The non-vanishing Dirac brackets are explicitly listed below

\begin{equation}
\label{f}
\lbrace\rho(x),\theta(y)\rbrace=\delta{\bf{(x-y)}};~ \lbrace\alpha(x),\theta(y)\rbrace=-\frac{\alpha}{\rho}\delta{\bf{(x-y)}};~\lbrace\alpha(x),\beta(y)\rbrace=\dfrac{\delta{\bf{(x-y)}}}{\rho}
.\end{equation}
Incidentally (\ref{f}) gives rise to two  independent canonical pairs  $(\rho , \theta ) $ and $(\alpha, \rho \beta )$.
 The canonical hamiltonian density  for the fluid corresponding to (\ref{c}) is,
\begin{equation}
\label{g}
{\cal{H}}=\pi_{\alpha}\dot{\alpha}+\pi_{\theta}\dot{\theta}+\pi_{\beta}\dot{\beta}+\pi_{\rho}\dot{\rho}-{\cal{L}}
$$$$= j^{i}\partial_{i}\theta+j^{i}\alpha\partial_{i}\beta+f(n).
\end{equation}
Using the Dirac brackets (\ref{f}) the hamiltonian equation of motion for $\rho$ is
\begin{equation}
\label{i}
\partial_{0} \rho=\lbrace \rho, H \rbrace~,~~~ H=\int{\cal{H}}d^{3}{\bf{x}},
\end{equation}
and we find
\begin{equation}
\label{k}
\dot \rho=-\partial_{i}j^{i},
\end{equation}
yielding the current conservation law (or in fluid dynamics terminology the continuity equation), obtained earlier (\ref{e0}).
In the same way we can find  equations of motion for $\alpha,\beta$,
\begin{equation}
\label{m}
\dot{\alpha}=\lbrace \alpha, H\rbrace;~~ \dot{\beta}=\lbrace \beta, H\rbrace
\end{equation}
from which we recover
\begin{equation}
\label{n}
\rho\dot{\alpha}=-j^{i}(\partial_{i}\alpha)\Rightarrow j^{\mu}\partial_{\mu}\alpha=0,
\end{equation}
and
\begin{equation}
\label{o}
 \rho\dot{\beta}=-j^{i}(\partial_{i}\beta)\Rightarrow j^{\mu}\partial_{\mu}\beta=0 .
\end{equation}
These equations are the same as the Euler-Lagrange  equations of motion  (\ref{e1},\ref{e2}). Finally, from $\dot{\theta}$ we find
\begin{equation}
\label{p}
\dot{\theta}=\lbrace \theta, H \rbrace =-\alpha\dot{\beta}-\dfrac{\rho}{n}{f^{\prime }(n)}.
\end{equation}
 This is same as (\ref{e3}) and equivalent to the time component of (\ref{r}).
In our case, the space components of (\ref{r}) just correspond to the equation for the nondynamical variable $j^{i}$.

At this point let us pause to note the status of the identity (\ref{r}). On one hand the $j_i$ variables are not involved in the symplectic structure (\ref{f}) and so should trivially commute with all degrees of freedom but on the other hand they are directly related to the dynamical variables through (\ref{r}) and infact yield non-zero brackets, {\it{e.g}}
$$\{j_i(x),\rho (y) \}=-\frac{n}{f'(n)}\{(\partial_i\theta +\alpha \partial _i\beta)(x),\rho (y)\}=\frac{n}{f'(n)}\partial_{i}\delta (x-y).$$

It is clear therefore that directly using $j_{i}$ or replacing it by the identity (\ref{r}) will yield distinct results in the calculation of brackets. This necessitates a specific prescription that will soon be elaborated.

 To illuminate the various issues let us now proceed to verify the Schwinger condition, which is a prerequisite for a relativistic field theory. Quite surprisingly, we will find that there are subtleties involved even in the free fluid theory and serious complications in the interacting theory of a fluid with external gauge field, to be treated in a later section. The problem is centered around the implementation of the space component of the relation (\ref{r}) and the construction of the symmetric energy-momentum (or stress) tensor $\Theta_{\mu \nu}$ required to formulate the Schwinger condition.

 The stress tensor is   obtained from $\cal{L}$ in a straightforward way \cite{jac}:
\begin{equation}
\label{s}
\Theta_{\mu \nu}=-\frac{2}{\sqrt{-g}}\frac{\partial S}{\partial g^{\mu\nu }}=-{\cal{L}}\eta_{\mu \nu}+ \dfrac{j_{\mu}j_{\nu}}{\sqrt{j^{2}}}f^{\prime}(\sqrt{j^{2}}).
\end{equation}
From (\ref{a}) and (\ref{r}) the above expression for the stress tensor can be written as,
\begin{equation}
\label{t}
\Theta_{\mu \nu}= -\eta_{\mu \nu}[nf^{\prime}(n)-f(n)]+\frac{j_{\mu}j_{\nu}}{n}f^{\prime}(n)
\end{equation}
which has the expected structure (\ref{bl}).
By comparison it is easy to obtain the identifications,
\begin{equation}
\label{bk}
 P_{rel}=nf'(n)-f(n), ~\epsilon_{rel}+P_{rel}=nf'(n),~j_{\mu}=nu_{\mu},
\end{equation}

The hamiltonian density from $\Theta_{\mu \nu}$ is given by,
\begin{equation}
\label{u}
\Theta_{00}=\frac{j_{i}j_{i}}{n}f^{\prime}(n)+f(n).
\end{equation}
To rewrite $\Theta_{00}$ in terms of Clebsch variables, we use (\ref{r})
\begin{equation}
\label{v}
j_{i}=-\frac{n}{f^{\prime}(n)}(\partial_{i}\theta + \alpha \partial_{i}\beta),
\end{equation}
and can recover the canonical form of the hamiltonian obtained earlier (\ref{g}), provided we replace only one of the $j^i$ in the quadratic term, leading to
\begin{equation}
\label{u1}
\Theta_{00}=j^{i}(\partial_{i}\theta + \alpha\partial_{i} \beta)+f(n).
\end{equation}
We stress that only this prescription will lead to the canonical  expression for the hamiltonian computed earlier, (that generated the correct dynamical equations). This is further corroborated by constructing the  momentum density,
\begin{equation}
\label{x}
\Theta_{0i}=\frac{j_{0}j_{i}}{n}f^{\prime}(n)=-\rho(\partial_{i}\theta+ \alpha\partial_{i}\beta),
\end{equation}
where, once again, the same prescription of replacing $j_{i}$ is exploited. It is straightforward to show that $\Theta_{0i}$ acts as the proper translation generator. Below we explicitly demonstrate this for $\alpha$:
\begin{equation}
\label{aa}
\lbrace\alpha, \int d\bar{x}\Theta_{0i}\rbrace \linebreak\\
=\lbrace \alpha, \int -\rho(\partial_{i}\theta+ \alpha\partial_{i}\beta) \rbrace \linebreak\\
=-(\partial_{i}\rho)\frac{\alpha}{\rho}+\frac{\partial_{i}(\rho \alpha)}{\rho}
=\partial_{i}\alpha.
\end{equation}
Likewise one may proceed for other variables.

It is important to note that, like $\Theta_{00}$, $\Theta_{0i}$ also agrees with the result obtained from the canonical stress tensor obtained via Noether prescription in (\ref{a}).
\begin{equation}
\label{newnoe}
T_{\mu\nu}=\frac{\partial {\cal{L}}}{\partial(\partial_{\mu}\theta)}\partial_{\nu}\theta + \frac{\partial {\cal{L}}}{\partial(\partial_{\mu}\beta)}\partial_{\nu}\beta +\frac{\partial {\cal{L}}}{\partial(\partial_{\mu}\alpha)}\partial_{\nu}\alpha +\frac{\partial {\cal{L}}}{\partial(\partial_{\mu}\rho)}\partial_{\nu}\rho - \eta_{\mu\nu}{\cal{L}}$$$$
=-j_{\mu}\partial_{\nu}\theta - \alpha j_{\mu}\partial_{\nu}\beta-\eta_{\mu\nu}{\cal{L}}.
\end{equation}
The $T_{0i}$ component is given by,
\begin{equation}
T_{0i}=-\rho \partial_{i}\theta-\alpha \rho \partial_{i}\beta
\end{equation}
which reproduces (\ref{x}).

Indeed, following our prescription of replacing $j_{\nu}$ in (\ref{s}) in favour of the Clebsch variables by exploiting(\ref{r}) immediately shows the exact equivalence between $\Theta_{\mu \nu}$(\ref{s}) and $T_{\mu \nu}$(\ref{newnoe}).

As is well known the definition of Noether charges may differ by local counter-terms. By appropriate manipulations it is however possible to abstract both $T_{\mu \nu}$ and $\Theta_{\mu \nu}$ from Noether's theorem \cite{zos}. However it must be  realised that in general $T_{\mu \nu}$ and $\Theta_{\mu\nu}$ are not identical. Indeed, by their very definitions (\ref{newnoe}) and (\ref{s}), respectively, it is seen that while $\Theta_{\mu\nu}$ is symmetric, $T_{\mu\nu}$ is not. For gauge theories the difference is proportional to the Gauss constraint so that $T_{\mu \nu}$ and $\Theta_{\mu \nu}$ agree on the physical subspace. The present theory is not a gauge theory as it is bereft of any first class constraint. Nevertheless we find that in the present case $T_{\mu \nu}$ and $\Theta_{\mu \nu}$ are identical provided we interpret $j_{\mu}$ in favour of Clebsch variables(\ref{r}), as already discussed. This interpretation is important and also plays a significant role in the derivation of the Schwinger condition discussed in the next subsection. In the interacting case to be considered in the  section 4, however, there is a difference between $T_{\mu \nu}$ and $\Theta_{\mu \nu}$ inspite of this particular interpretation of $j_{\mu}$. But, by improving $T_{\mu \nu}$ (which is similar to Belinfante's prescription), it becomes identical to $\Theta_{\mu \nu}$.

\subsection{Conservation laws in the hamiltonian formulation and Schwinger condition}

The analysis of fluids done here strongly rests on the conservation laws (\ref{one}) and (\ref{e0}) for the stress tensor and current, respectively. It would be worthwhile to obtain these relations in a hamiltonian approach. That would also clarify the role and utility of the Schwinger condition.

Let us begin by considering the algebra of $\Theta _{00}$ with $j_{0}$, 
\begin{equation}
\label{j0}
\{j_{0}(x), \Theta_{00}(y)\}=\{j_{0}(x), j^{i}(\partial_{i}\theta + \alpha\partial_{i} \beta)(y)+f(n)(y)\}
\end{equation}
The only nontrivial bracket of $j_{0}$(or $\rho$) is with the $\theta$ variable. Using (\ref{f}) we obtain,
\begin{equation}
\label{w0}
\{j_{0}(x), \Theta_{00}(y)\}=j^{i}(y)\partial^{y}_{i}\delta(x-y)
\end{equation}
which reproduces the expected algebra. Its integrated version immediately yields (\ref{e0}). To see this consider the above algebra by integrating over $y$,
\begin{equation}
\label{w1}
\{j_{0}(x), \int d^{3}y \Theta_{00}(y)\}=\int d^{3}y j^{i}(y)\partial^{y}_{i}\delta(x-y)
\end{equation}
Recalling that $\int d^{3}y \Theta_{00}(y)$ is the hamiltonian we obtain, by dropping a surface term, 
\begin{equation}
\label{w2}
\partial_{0}j_{0}=-\partial_{i}j^{i}
\end{equation}
thereby reproducing (\ref{e0}).

We now consider the algebra of $\Theta_{00}$ with itself. This algebra is the famous Schwinger condition whose integrated version would yield (\ref{one}), similar to the above derivation of (\ref{e0}).
\begin{equation}
\label{ac}
\lbrace \Theta_{00}({x}),\Theta_{00}({y})\rbrace=\lbrace j^{i}(\partial_{i}\theta + \alpha\partial_{i} \beta)(x)+f(n)(x), j^{k}(\partial_{k}\theta + \alpha\partial_{k} \beta)(y)+f(n)(y)\rbrace .
\end{equation}
Exploiting the basic brackets (\ref{f}) we find,
\begin{equation}
\{\Theta_{00}({x}),\Theta_{00}({y})\}=\bigl [\frac{j_{i}(x)f'(x) \rho(x)}{n(x)}+\frac{j_{i}(y)f'(y) \rho(y)}{n(y)}\bigr ]\partial_{i}^{x}\delta{\bf{(x-y)}}.
\end{equation}
Recalling the identification of $j_{i}$ in terms of the Clebsch variables (\ref{v}) we obtain,
\begin{equation}
\label{ad}
\lbrace \Theta_{00}(x),\Theta_{00}(y)\rbrace= - \bigl[(\rho(\partial_{i}\theta + \alpha\partial_{i} \beta)(x)+\rho(\partial_{i}\theta + \alpha\partial_{i} \beta)(y)\bigr]\partial_{i}^{x}\delta{\bf{(x-y)}}.
\end{equation}
The expression on the right side is now expressed in terms of $\Theta_{0i}$ by using (\ref{x})
\begin{equation}
\label{ab}
\lbrace \Theta_{00}(x),\Theta_{00}(y)\rbrace=(\Theta_{0i}(x)+\Theta_{0i}(y))\partial^{(x)}_{i}\delta{\bf{(x-y)}},
\end{equation}
which is the Schwinger condition\cite{nger}. 

Let us now consider its integrated version, 
\begin{equation}
\label{sci}
\lbrace \Theta_{00}(x),\int d^{3}y \Theta_{00}(y)\rbrace = \int d^{3}y (\Theta_{0i}(x)+\Theta_{0i}(y))\partial^{(x)}_{i}\delta{\bf{(x-y)}}
\end{equation}
which simplifies, after dropping surface terms, as, 
\begin{equation}
\partial_{0}\Theta_{00}=\partial_{i}\Theta_{0i}
\end{equation}
which is just the time component of (\ref{one})
\begin{equation}
\label{consrv}
\partial_{\mu}\Theta^{\mu 0}=0
\end{equation}

Likewise the space component of (\ref{one}) may be obtained from other Schwinger conditions that involve the algebra among $\Theta_{00}-\Theta_{0i}$ and $\Theta_{0i}-\Theta_{0j}$. It is useful to mention that, at an intermediate stage, we have to use the relation,
\begin{equation}
u^{\mu}(\partial_{\nu}u_{\mu}-\partial_{\mu}u_{\nu})f'+(g_{\mu\nu}-u_{\mu}u_{\nu})\partial^{\mu}nf''=0.
\label{rel}
\end{equation}
which may also be verified explicitly. This is the relativistic generalization of the  Euler equation as noted by \cite{jac}. Although in non-relativistic fluid mechanics, Euler equation is frequently used, quite surprisingly the relativistic Euler equation is not very familiar. 

It is perhaps pertinent to mention that the Schwinger condition was originally proposed in the context of relativistic quantum field theory. This was an alternative route to establish the conservation of the stress tensor as well as the validity of the Poincare algebra. Nevertheless, it has also found applications in discussing analogous features in the context of classical field theory\cite{ra, rasu}. The point is that while the validity of the Schwinger condition is not mandatory in the classical context, any deviation must be such that the integrated version leads to the conservation law (\ref{one}). In the present case we find that the Schwinger condition holds exactly. This is a new finding in the context of classical fluids.

It is useful to recall that the Schwinger condition was derived for the symmetric stress tensor $\Theta_{\mu\nu}$ defined in (\ref{s}). Since the proof relies on this symmetricity  it does not, in general, hold for $T_{\mu\nu}$ defined in (\ref{newnoe}). The nice point of our analysis is that, subject to the interpretation of $j_{\mu}$ discussed previously, it is possible to recast $T_{\mu\nu}$ in a symmetric form that is identical to $\Theta_{\mu\nu}$. This appears to be a unique characteristic of the theory of classical fluids developed here.
 There are important physical implications  of the Schwinger condition for classical fluids. The first point to note is that the conservation 
law (\ref{one}) is the fundamental equation on which the dynamics of fluids is based. Establishing Schwinger condition automatically implies (\ref{one}). Next, the role of Clebsch variables gets illuminated. As discussed previously, one  of the $j_{i}$ in $\Theta_{00}$(\ref{u}) has to be eliminated in favour of these variables to get (\ref{u1}) which reproduces the equations of motion for the basic variables. It is now found  that exploiting precisely this structure of $\Theta_{00}$, the Schwinger condition holds. This serves as an important consistency check on our formalism. As a side remark we find that  the same prescription also leads to current conservation(\ref{e0}) starting from the algebra (\ref{w0}).
\vspace{1cm}

\section{Relativistic fluid mechanics in light-cone (null plane) coordinates:}
In this section we study fluid mechanics in light-cone coordinates. Apart from providing a different formulation than the equal time one, there is another motivation which will become clearer in the next section when we discuss the non-relativistic reduction of the fluid model. We define the light-cone coordinates   as in \cite{hrt},
  $\lbrace x^{+}, x^{-}, \bar{x}\rbrace$ where $\bar{x}\equiv x^1,x^2$ and $x^{\pm}=\frac{1}{\sqrt 2}(x^0\pm x^3)$. The nonvanishing  metric components are   $g^{+-}=g^{-+}=1,~g^{ii}=-1, i,j=1,2$.  The fluid lagrangian  in this coordinate system is,
 \begin{equation}
 \label{af}
{\cal{L}}=-j^\mu a_\mu -f(\sqrt{j^\mu j_\mu })=-(j^+a_++j^-a_- +j^ia_i)-f =-(j_-a_++j_+a_- -j_ia_i)-f $$$$
=-j_+(\partial _-\theta +\alpha \partial _-\beta )-j_-(\partial _+\theta +\alpha \partial _+\beta )+j_ia_i-f,\\
\end{equation}\\
where, in the last step,  we have exploited the definition of $a_\mu $ (\ref{b}).
Note that $x^+$ plays the role of  time and the dynamical variables are identified following our previous prescription, that is variables involved in $x^+$-derivatives only are considered as dynamical. In the present setup the degrees of freedom are $j_-, \theta, \alpha, \beta $.
The  momentum is defined as
\begin{equation}
\pi _\phi =(\partial L)/(\partial (\partial _+\phi ))
\label{ag}
\end{equation}
for a generic $\phi $ and  $\partial_{+}\equiv\partial_{t}.$
The first order model (\ref{af})  produces the constraints,
\begin{equation}
\chi_1=\pi_\theta +\rho \approx 0~,~\chi_2=\pi_\beta +\rho\alpha\approx 0 ~,~\chi_3=\pi_\alpha\approx 0 ~,~\chi_4=\pi^{-}\approx 0 .
\label{ah}
\end{equation}
where $\pi^{-}$is the momenta conjugate to $j_{-}$. Note that $j_{-}$ has to be identified with $\rho$.
Constraint analysis once again provides the Dirac brackets
\begin{equation}
\{\rho (x),\theta (y)\}=\delta {\bf{(x-y)}}~,~\{\alpha (x),\theta (y)\}=-(\alpha /\rho )\delta {\bf{(x-y)}}~,~\{\alpha (x),\beta(y)\}=(1/\rho )\delta {\bf{(x-y)}},
\label{ai}
\end{equation}
where ${\bf{x}}=x^{-}, \bar{x}$ with $ \bar{x}=x^1,x^2 $ and $\delta {\bf(x-y)}=\delta (x^--y^-)\delta (\bar{x}-\bar{y})$.
It is worthwhile to point out that the above bracket structure in light cone coordinates is same as the one derived earlier in (\ref{f}) in equal time  coordinate system. This is simply because the lagrangian (\ref{c}) was also first order. {\footnote{This can be contrasted with a generic second order system, {\it{e.g.}} Klein-Gordon lagrangian, whose light-cone reduction yields a first order system with a drastically altered constraint structure.}
The hamiltonian density is given by
$${\cal{H}}=\pi_{\alpha}\dot{\alpha}+\pi_{\theta}\dot{\theta}+\pi_{\beta}\dot{\beta}+\pi_{-}\dot{j_{-}}-{\cal{L}},$$
from which, using (\ref{af}) and (\ref{ah}), the hamiltonian of the fluid is,
   \begin{equation}
H=\int dx^- d\bar {x}~{\cal{H}}(x)=\int dx^- d\bar x~[j_+(\partial _-\theta +\alpha \partial _-\beta )-j_ia_i+f].
\label{aj}
\end{equation}
Before proceeding further we need to check the overall consistency of the light-cone framework mainly because of our specific interpretation of the space component of (\ref{r}) and its subsequent applications.

Let us start by comparing the lagrangian and hamiltonian equations of motion. First comes the continuity equation. From the lagrangian (\ref{af})  by varying $\theta$ we obtain,
\begin{equation}
\label{th}
\partial _+j_{-}  +  \partial _-j_+  -\partial _ij_i =\partial _\mu j^\mu =0
\end{equation}
which is the continuity equation in light-cone coordinates. On the other hand, in the hamiltonian framework, we have
\begin{equation}
\label{con}
\partial_{+}j_{-} (x)=\lbrace j_{-}(x),H\rbrace =\lbrace j_-(x),\int dy^-d\bar y~(j_+(\partial _-\theta +\alpha \partial _-\beta )-j_ia_i+f)\rbrace $$$$
=-\partial_{-}j_{+}(x)+\partial_{i}j_{i}(x),
\end{equation}
which reproduces (\ref{th}). It is interesting to observe that the spatial part is now broken up into two sectors $x^-$ and $\bar{x}$ that are qualitatively somewhat distinct.

Let us rederive the light-cone version of the rest of the lagrangian variational equations (\ref{e1}-\ref{e3}). The hamiltonian equation,
\begin{equation}
\partial_{+}\alpha=\lbrace \alpha(x) ,H\rbrace
=-\frac{(\partial_{-}\alpha) j_{+}}{j_{-}}+\frac{(\partial_{i}\alpha) j_{i}}{j_{-}},
\end{equation}
can be rearranged to yield (\ref{e1}) while
\begin{equation}
\partial_{+}\beta = \lbrace \beta (x),H\rbrace
 =-\frac{(\partial_{-}\beta) j_{+}}{j_{-}}+\frac{(\partial_{i}\beta) j_{i}}{j_{-}},
\end{equation}
reproduces (\ref{e2}). In a similar way  $\partial_{+}\theta$ obtained below
\begin{equation}
\label{an}
\partial_{+}\theta=  \lbrace \theta (x) ,H\rbrace
=\frac{\alpha j_{+}(\partial_{-}\beta)}{j_{-}} - \frac{\alpha j_{i}(\partial_{i}\beta)}{j_{-}}-\frac{f^{\prime}j_{+}}{n}
\end{equation}
is the light-cone version of (\ref{e3}).
\subsection{Conservation laws in hamiltonian formulation and Schwinger condition}

In order to discuss the conservation laws in the light-cone coordinates we have to first identify the appropriate hamiltonian. Consider the $\Theta_{+-}$ component of (\ref{t}),
\begin{equation}
\nonumber
\Theta_{+-}=-(nf'-f)g_{+-}+\frac{f'}{n}j_{+}j_{-}
\end{equation}
\begin{equation}
\nonumber
=f-\frac{f'}{n}(j_{+}j_{-}-j_{i}j_{i})
\end{equation}
\begin{equation}
=f+j_{+}a_{-}-j_{i}a_{i}.
\end{equation}
We identify this with the canonical hamiltonian density $({\cal{H}})$ defined in (\ref{aj}). This may be easily seen by replacing $a_{-}$ using (\ref{b}).

We are now ready to obtain the various conservation laws. Let us first derive the result (\ref{th}). This will also act as a forerunner for the derivation of the Schwinger condition in light cone coordinates. Consider the algebra,
\begin{equation}
\label{con1}
\lbrace j_{-}(x),\Theta_{+-}(y)\rbrace = \{j_{-}(x),(j_{+}a_{-}-j_ia_i+f)(y)\}
\end{equation}
Replacing $a_{-}$ and $a_{i}$ from (\ref{b}) and using the algebra (\ref{ai}) yields,
\begin{equation}
\label{j3}
\lbrace j_{-}(x),\Theta_{+-}(y)\rbrace = j_{+}(y)\partial^{y}_{-}\delta(x-y)-j_{i}(y)\partial^{y}_{i}\delta(x-y).
\end{equation}
Taking its integrated version,
\begin{equation}
\label{j4}
\lbrace j_{-}(x),\int d^{3}y ~\Theta_{+-}(y)\rbrace =\int d^{3}y ~(j_{+}(y)\partial^{y}_{-}\delta(x-y)-j_{i}(y)\partial^{y}_{i}\delta(x-y)).
\end{equation}
and dropping the surface terms yields,
\begin{equation}
\label{con2}
\partial_{+}j_{-} (x)=-\partial_{-}j_{+}(x)+\partial_{i}j_{i}(x),
\end{equation}
which reproduces (\ref{th}).

We next discuss the Schwinger condition. The relevant algebra is,
\begin{equation}
\label{sch}
\{\Theta_{+-}(x),\Theta_{+-}(y)\}=\{[j_+(\partial _-\theta +\alpha \partial _-\beta )-j_ia_i+f](x), [j_+(\partial _-\theta +\alpha \partial _-\beta )-j_ia_i+f](y)\}
\end{equation}
 After some algebra we end up with,
\begin{equation}
\label{sch1}
\{\Theta_{+-}(x),\Theta_{+-}(y)\}=-j_+(x)\partial_-^x(\frac{f'j_+\delta {\bf{(x-y)}}}{n})+j_+(y)\partial_-^y(\frac{f'j_+\delta {\bf{(x-y)}}}{n})$$$$
+j_i(x)\partial_i^x(\frac{f'j_+\delta {\bf{(x-y)}}}{n})-j_i(y)\partial_i^y(\frac{f'j_+\delta {\bf{(x-y)}}}{n}).
\end{equation}
On further simplification we obtain,
\begin{equation}
\label{sc2}
\{\Theta_{+-}(x),\Theta_{+-}(y)\}=\bigl[\frac{f'(j_+)^2}{n}(x)+\frac{f'(j_+)^2}{n}(y)\bigr]\partial^{y}_-\delta {\bf{(x-y)}}+ \bigl[\frac{f'j_+j_i}{n}(x)+\frac{f'j_+j_i}{n}(y)\bigr]\partial^{x}_i\delta{\bf{(x-y)}} .
\end{equation}
From (\ref{r}) and (\ref{t}) we identify the other components of the stress tensor, $$ \frac{f'(j_+)^2}{n}= \Theta_{++}~,~\frac{f'j_+j_i}{n}=\Theta_{+i}~,$$ and thereby recover  the cherished form of the  Schwinger condition in light-cone coordinates,
\begin{equation}
\label{sc3}
\{\Theta_{+-}(x),\Theta_{+-}(y)\}= -(\Theta_{++}(x)+\Theta_{++}(y))\partial_-\delta {\bf{(x-y)}}+(\Theta_{+i}(x)+\Theta_{+i}(y))\partial_i\delta {\bf{(x-y)}}.
\end{equation}
We emphasize that this is a completely new result in the context of light-cone formulation of classical fluid.

Integrating over $y$ we recover
\begin{equation}
\label{con1}
\partial_+\Theta_{+-} =-\partial_-\Theta_{++}+\partial_i\Theta_{+i}
\end{equation}
or equivalently the energy conservation condition
\begin{equation}
\label{}
\partial_+\Theta^{+-} +\partial_-\Theta^{--}+\partial_i\Theta^{i-}=0
\end{equation}
since this is the $\nu =-$ component of the conservation law(\ref{one}). Note that this computation can be repeated for $\nu =+,i$ but infact that is unnecessary since the covariant conservation law follows directly from the lagrangian (\ref{f}) and we have checked individually that the hamiltonian equations of motion in light-cone coordinates match correctly with their lagrangian counterpart. Finally, as discussed in Section 2, the light-cone version of the relativistic Euler equation(\ref{rel}) will also appear in the present setup.

To the best of our knowledge, in  our work, for the first time, the light-cone analysis of relativistic fluid model has been carried through where  the specific identification of the physical degrees of freedom with the Clebsch variables  has been spelt out.

\section{Interacting fluid model} 
The background gauge field $A_{\mu}$ is introduced in the fluid lagrangian in a conventional way,
 \begin{equation}
\label{newa}
 {\cal{L}}=-\eta^{\mu\nu}j_{\mu}(a_{\nu}-A_{\nu})-f.
 \end{equation}
 
Here also $j^{i}$ is regarded as an auxiliary  variable. The dynamical equations which are modified by the gauge field are provided below,
      \begin{equation}
     \label{newe1}
     \dot{\theta}+ \alpha\dot{\beta}+\dfrac{\rho}{n}{f^{\prime }(n)}-A_{0}=0.
  \end{equation}
  \begin{equation}
\label{newr}
j_{\mu}=-\frac{n}{f^{\prime}(n)}(a_{\mu}-A_{\mu})=-\frac{n}{f^{\prime}(n)}(\partial_{\mu}\theta + \alpha \partial_{\mu}\beta -A_{\mu}).
\end{equation}
Rest of the equations of motion are same as the free theory, given in (\ref{e1}, \ref{e2}). Notice that the conjugate momenta remain unaffected (\ref{d}) since no new time-derivatives are introduced in the interacting theory and hence the same Dirac bracket structure (as in the free fluid theory) will prevail.

 The canonical Hamiltonian is given by
\begin{equation}
\label{newg}
{\cal{H}}=\pi_{\alpha}\dot{\alpha}+\pi_{\theta}\dot{\theta}+\pi_{\beta}\dot{\beta}+\pi_{\rho}\dot{\rho}-{\cal{L}}
$$$$= j^{i}\partial_{i}\theta+j^{i}\alpha\partial_{i}\beta - j^{\mu}A_{\mu} +f(n).
\label{nh}
\end{equation}
 The $\theta $ equation is recovered below,
\begin{equation}
\label{newp}
\dot{\theta}=\lbrace \theta, H \rbrace =-\alpha\dot{\beta}-\dfrac{\rho}{n}{f^{\prime }(n)}+A_{0}.
\end{equation}
Rest of the equations of motion are also derived correctly. Thus the hamiltonian in  (\ref{nh}) is able to generate the correct dynamics.

Following our free theory analysis we now derive the  covariant stress tensor  $\Theta_{\mu \nu}$ for the interacting theory,
\begin{equation}
\label{news}
\Theta_{\mu \nu}=-\frac{2}{\sqrt{-g}}\frac{\partial S}{\partial g^{\mu\nu }}=-{\cal{L}}\eta_{\mu \nu}+ \dfrac{j_{\mu}j_{\nu}}{\sqrt{j^{2}}}f^{\prime}(\sqrt{j^{2}})$$$$
= -(-j^{\sigma}(a_{\sigma}-A_{\sigma})-f)\eta_{\mu \nu}+ \dfrac{j_{\mu}j_{\nu}}{\sqrt{j^{2}}}f^{\prime}(\sqrt{j^{2}}).
\end{equation}
We express $\Theta_{\mu \nu}$ in terms of Clebsch variables following our earlier prescription of replacing $j_{\nu}$ by exploiting (\ref{newr}), 
\begin{equation}
\label{news1}
\Theta_{\mu \nu}= -(-j^{\sigma}(a_{\sigma}-A_{\sigma})-f)\eta_{\mu \nu}- j_{\mu}(\partial _\nu\theta +\alpha \partial _\nu \beta-A_{\nu}).
\end{equation}
One can directly check that $\Theta_{\mu \nu}$ satisfies the correct conservation law in presence of interactions,
\begin{equation}
\nonumber
\partial^{\mu}\Theta_{\mu\nu}=-\partial_{\nu}[-j^{\mu}(a_{\mu}-A_{\mu})-f]-j_{\mu}\partial^{\mu}[\partial_{\nu}\theta+\alpha\partial_{\nu}\beta-A_{\nu}]
\end{equation}
\begin{eqnarray}
 \nonumber
=\partial_{\nu}j^{\mu}(\partial_{\mu}\theta + \alpha \partial_{\mu}\beta -A_{\mu})+j^{\mu}\partial_{\nu}(\partial_{\mu}\theta + \alpha \partial_{\mu}\beta -A_{\mu})+\partial{\nu}f \\
- j_{\mu}\partial^{\mu}\partial_{\nu}\theta-\alpha j_{\mu}\partial^{\mu}\partial_{\nu}\beta+ j_{\mu}\partial^{\mu}A_{\nu}
\end{eqnarray}
\begin{equation}
=j^{\mu}F_{\mu\nu}+\partial_{\nu}f+\partial_{\nu}j^{\mu}(\partial_{\mu}\theta + \alpha \partial_{\mu}\beta -A_{\mu})= j^{\mu}F_{\mu\nu}.
\label{hu}
\end{equation}

where we have exploited the result (\ref{newr}).
The hamiltonian density obtained from (\ref{news1}) is given by,
\begin{equation}
\label{newss}
\Theta_{00}=j^{i}(a_{i}-A_{i})+f=j^{i}(\partial_{i}\theta +\alpha \partial_{i}\beta -A_{i})+f.
\end{equation}
Immediately we are faced with a problem: the expressions for the  hamiltonian density given in (\ref{newg}) and (\ref{newss}) do not match. The mismatch term is $j_{0}A_{0}$ which has nontrivial brackets with $\theta$. Thus the hamiltonian density (\ref{newss}) fails to generate the lagrangian equation of motion for the $\theta $ variable (\ref{newe1}).  Of course in the absence of interaction the results agree.

The expression for the canonical stress tensor $T_{\mu \nu}$ is straightforward to obtain following the Noether prescription. The result is (\ref{newnoe}) with the lagrangian ${\cal{L}}$ defined in (\ref{newa}). Obviously $T_{00}$ agrees with the canonical hamiltonian density (\ref{nh}). Also $T_{0i}$ following from (\ref{newnoe}) and (\ref{newa}),$$T_{0i}= \pi_{\theta}\partial_{i}\theta+ \pi_{\beta}\partial_{i}\beta=-\rho(\partial_{i}\theta + \alpha \partial_{i}\beta).$$ matches with the non-interacting fluid result (\ref{x}), and behaves like the correct translation generator. In obtaining the final expression we have imposed the constraints (\ref{e}) strongly since Dirac brackets are being ussed. Using (\ref{f}) we obtain,
\begin{equation}
\label{aaaaa}
\lbrace \theta, \int d\bar{x} T_{0i}\rbrace \linebreak\\
=\lbrace \theta, \int -\rho(\partial_{i}\theta+ \alpha\partial_{i}\beta) \rbrace \linebreak\\
=\partial_{i}\theta
\end{equation}
which is the desired translation law.

 However, $\Theta_{0i}$ defined from (\ref{news1}),$$\Theta_{0i}= -\rho(\partial_{i}\theta+\alpha \partial_{i}\beta-A_{i}),$$ does not match with $T_{0i}$, and it does not correctly generate the translation of $\theta$,
\begin{equation}
\label{aaaa}
\lbrace\theta, \int d\bar{x}\Theta_{0i}\rbrace \linebreak\\
=\lbrace \theta, \int -\rho(\partial_{i}\theta+ \alpha\partial_{i}\beta-A_{i}) \rbrace \linebreak\\
=\partial_{i}\theta+A_{i}.
\end{equation}

 Let us next derive the conservation law satisfied by $T_{\mu\nu}$.  Taking a four-divergence of (\ref{newnoe})
yields,
\begin{equation}
\nonumber
\partial^{\mu}T_{\mu\nu}=-\partial^{\mu}(j_{\mu}\partial_{\nu}\theta)-\partial^{\mu}(\alpha j_{\mu}\partial_{\nu}\beta) -\partial_{\nu}{\cal{L}}.
\end{equation}
Exploiting the equations of motion we find
\begin{equation}
\partial^{\mu}T_{\mu\nu}=(\partial_{\nu}j^{\mu})\partial_{\mu}\theta-j_{\mu}\alpha \partial^{\mu}\partial_{\nu}\beta -(\partial_{\nu}j^{\mu})A_{\mu}+\partial_{\nu}f $$$$
=j^{\mu}F_{\mu\nu}-j^{\mu}\partial_{\mu}A_{\nu}-(\partial_{\nu}j^{\mu})A_{\mu}+(\partial_{\nu}j^{\mu})\partial_{\mu}\theta + \alpha(\partial_{\nu}j^{\mu})\partial_{\mu}\beta  +\partial_{\nu}f $$$$
=j^{\mu}F_{\mu\nu}-\partial_{\mu}(j^{\mu}A_{\nu}).
\end{equation}
First of all, in the absence of $A_\mu $ the stress tensor is conserved. This is compatible with the free fluid theory discussed in section 2. But for the interacting theory the stress tensor does not reproduce  the expected conservation law, as computed in (\ref{hu}). Apart from the Lorentz force term there is an additional piece. However it is possible to define an 'improved' canonical stress tensor $\tilde{T_{\mu \nu}}$ that yields the desired relation. It is given by,
\begin{equation}
\label{imp}
\tilde{T_{\mu \nu}}=T_{\mu \nu}+ j_{\mu}A_{\nu}
\end{equation}
which satisfies,
\begin{equation}
\partial^{\mu} \tilde{T_{\mu \nu}}=j^{\mu} F_{\mu \nu}
\label{conser1}
\end{equation}

It is now possible to show that this $\tilde{T_{\mu\nu}}$ is exactly identical to $\Theta_{\mu\nu}$ (\ref{news}). From
(\ref{newnoe}) and (\ref{imp}) we obtain
\begin{equation}
\label{til}
\tilde{T_{\mu \nu}}=-j_{\mu}(\partial_{\nu} \theta +\alpha \partial_{\nu} \beta- A_{\nu})-\eta_{\mu \nu}{\cal{L}}
\end{equation}
Exploiting (\ref{newr}) we find,
\begin{equation}
\tilde{T_{\mu\nu}}=-{\cal{L}}\eta_{\mu \nu}+ \dfrac{j_{\mu}j_{\nu}}{\sqrt{j^{2}}}f^{\prime}(\sqrt{j^{2}})
\end{equation}
which is the same as $\Theta_{\mu\nu}$ defined in (\ref{news}).

It is worthwhile to observe the complementary roles of the canonical (Noether) stress tensor $(T_{\mu\nu})$ and the symmetric (Schwinger) stress tensor $(\Theta_{\mu\nu})$. While the canonical expression correctly reproduces the equations of motion for all the dynamical variables, the symmetric one fails for the $\theta$ variable. On the other hand the symmetric tensor yields the correct Lorentz force term but the canonical tensor fails. Nevertheless, it is possible to redefine the latter from the conservation law such that the expected result is reproduced. Furthermore, this 'improved' canonical tensor matches exactly with the symmetric one.
\vspace{1cm}

\section{Conclusion and future prospects}
Fluid dynamics has generally been considered as an applied science but there has been a paradigm shift in modern physics perspective where deep theoretical aspects of the theory are being studied in the context of Fluid/Gravity correspondence \cite{ranga}, conformal symmetry of non-relativistic fluid dynamics \cite{gold, jl}, etc. A stepping stone in this direction would be to study fluid dynamics from a modern field theory point of view. This has been the motivation of the present paper. We have discussed kinematic and dynamic aspects  in detail both for ideal and interacting fluids, the latter  in the presence of gauge fields. We have principally used a hamiltonian formalism since this framework is most appropriate for studying symmetry properties. The Clebsch parametrization plays an essential role in our framework where the fluid turns out to be a second class constraint system. We have reconsidered the fluid model in light-cone coordinate system which is  qualitatively different from the equal-time coordinate system considered earlier. The light-cone analysis of fluids has recently attracted a lot of attention \cite{son}.
 
 In both equal-time as well as light-cone formulation we have shown the validity of the Schwinger condition, a hallmark of any relativistic field theory. Although the Schwinger condition was orginally given for relativistic quantum field theory, there are instances\cite{ra, rasu} where it holds for the classical case also. We find here that it is valid for relativistic classical fluids. The Schwinger condition involves the computation of the algebra of the stress tensor components. Since the fluid is a constrained system, it is essential to use Dirac brackets to calculate this algebra. It needs to be emphasized that this computation is by no means straightforward and requires subtle interpretation of the auxiliary variables in terms of physical fluid degrees of freedom. This interpretation is completely new and was instrumental in our derivation of the Schwinger condition.
 
 The role of the Schwinger condition vis-a-vis the Clebsch parametrisation was highlighted. The utility of this parametrisation which is frequently used in the analysis of fluids\cite{baz, jac}, is manifested in the present case through the study of Schwinger condition.
 
 Another thrust of our work is in the study of fluids in the presence of external gauge interactions. We have demonstrated that the canonical (Noether) and symmetric forms of stress tensors do not match although both have essential properties pertaining to it such as generating proper dynamics (in case of the canonical one) and satisfying correct conservation principle (in case of the symmetric one). In this sense the two definitions of the stress tensor complement each other. However,  it still needs to be seen how to define a stress tensor that obeys both these properties. We have also shown how an elegant modification of the canonical stress tensor leads to the symmetric one. In this analysis we have once again used the same interpretation of the auxiliary variable in terms of the physical ones as done for the free theory. This shows the robustness of our interpretation.

There are diverse channels along which further work can be pursued. It will  be worthwhile to generalize our analysis for viscous fluids. Another open problem is the hamiltonian analysis of fluid interacting with dynamical gauge fields. Obviously this is a non-trivial extension where new symplectic structures will emerge. Moreover the energy density of the fluid discussed here is a function of $n$ only which corresponds to the barotropic fluid. But in general it is possible to have a dependence of the entropy density.  We hope to report on these findings in the near future.

\end{document}